# On the estimation of the evolutionary power spectral density

H.P. Hong
Email: hongh@eng.uwo.ca

Department of Civil and Environmental Engineering, University of Western Ontario,
London, Canada, N6A 5B9

**Abstract**: Two popular spectral-based approaches for estimating the evolutionary power spectral density (EPSD) function from the samples of the evolutionary process are based on the short-time Fourier transform (STFT) and the continuous wavelet transform.  Both rely on the concept of slowly varying modulation or EPSD function, although the quantification of the effect of the ''slow'' variation in the estimated EPSD is elusive.

We propose, in the present study, to use the derivatives of the EPSD function to quantify the smoothness of the EPSD function in the context of estimating the EPSD function.  We derive equations for estimating EPSD by using the S-transform and continuous wavelet transform.  These equations are as simple to use as that derived based on STFT.  We also derive the corresponding equations for assessing the residual for the estimated EPSD by using these transforms, including STFT.  The residual provides an approach for identifying or quantifying, in the context of its estimation, the ''slow'' variation of the EPSD function.  The derived equations and numerical results indicate that the residual depends on both the derivatives of the EPSD function with respect to time and frequency as well as the adopted transform.

(*Submitted to a Journal*)

## 1.0   Introduction

Two of the important structural design loads for buildings and infrastructure systems are the wind and earthquake loads.  The wind load depends on the time history of the wind speed and the earthquake load is characterized by the seismic ground motions.  The wind speed of high-intensity wind events (e.g., thunderstorms) and ground motions of large earthquakes (Newmark and Rosenblueth 1974; Simiu and Scanlan 1996) are stochastic and nonstationary.  Besides the wind speed and seismic ground motions, the natural and man-made materials could also be treated as spatially inhomogeneous/nonstationary stochastic processes/fields (Vanmarcke 2010; Phoon 2008).  The mentioned random phenomena are often modeled using the evolutionary stochastic process (Priestley 1965).

The evolutionary process is built upon the Cramer spectral decomposition of a stationary stochastic process (Cramer 1942) by including a modulation function (Priestley 1965), which is slowly varying.  The modulation function can be a function of time and frequency.  The process is known as the uniformly modulated evolutionary process if the modulation function is only time-varying.  The evolutionary power spectral density (EPSD) function for the evolutionary process is directly proportional to the square of the amplitude modulation function.  For many engineering applications (e.g., in modelling seismic ground motions and wind speed of high-intensity wind events), the assignment of the amplitude modulation function, hence, the EPSD function is based on the observed samples of the considered nonstationary process. Two well-known approaches for estimating the EPSD function were suggested in the literature. One was proposed by Priestley (1965) and the other was proposed by Spanos and Failla (2004). In the approach proposed by Priestley, the EPSD function for a given sample or time-varying signal of a stochastic process is estimated by first applying the short-time Fourier transform (STFT) to decompose the signal and to obtain the time-frequency dependent power distribution. This obtained power distribution is then weighted using a moving window along the time axis to reduce the sampling fluctuation. The approach proposed by Spanos and Failla (2004) was formulated by directly applying the continuous wavelet transform (CWT) to the evolutionary process and considering that the amplitude modulation function is slowly varying. However, since "it is difficult to quantify what 'slow' variation is" (Spanos et al. 2005), the impact of a quantifiable slow variation on the accuracy of the estimated EPSD function is not addressed in the literature.

Both of these approaches have been employed for solving practical engineering problems. For example, STFT based approach was employed to estimate the EPSD functions of the seismic ground motions (Sabetta and Pugliese 1996) and wind speed of thunderstorm winds (rear-flank downdraft) (Chen and Letchford 2005). The wavelet-based approach was used to estimate the EPSD function of the seismic ground motions (Spanos and Failla 2004; Spanos et al. 2005; Giaralis and Spanos2009; Huang and Chen 2009; Sarkar et al. 2016) and the wind speed (Huang and Chen 2009).

In addition to the above, the use of the S-transform (ST) (Stockwell et al. 1996) to estimate PSD function in the time-frequency domain for seismic ground motions and wind speed was also considered (Hong and Cui 2020; Huang et al. 2020; Cui and Hong 2021a; Hong et al 2021b; Hong and Cui 2023). These studies indicate that the estimated PSD function in the time-frequency domain by applying ST could be interpreted as an approximation to the EPSD function. Again, the impact of a potential quantifiable slow variation on the accuracy of this interpretation is unknown. It is worth noting that one could simulate the nonstationary non-Gaussian process based on the iterative power and amplitude correction (IPAC) algorithm (Hong et al. 2021a,c; Hong and Cui 2023) by using a selected transform pair and without invoking the concept and definition of the evolutionary process, where the PSD function is defined in the transform domain. The IPAC algorithm can also be used to generate samples of the multivariate nonstationary Gaussian/non-Gaussian processes (Cui and Hong 2021b). However, the IPAC algorithm for simulating the Gaussian process is computationally less efficient than the spectral representation method (Shinozuka and Jan 1972; Shinozuka and Deodatis 1991); the relation between the estimated EPSD function and the time-frequency dependent PSD function defined in the transform domain for a selected transform has not been clarified entirely.

In the present study, we propose to use the derivatives of the EPSD function as measures to quantify the smoothness of the EPSD function in the context of estimating the EPSD function and to assess the residual of the estimate. The considered transforms for estimating the EPSD function are the STFT with the box window and Gaussian window, (generalized) ST, and CWT with the harmonic wavelet and the generalized Morse wavelet. Simple equations for estimating the EPSD function and the corresponding residual or error term for the estimation are derived and presented for each of the considered transforms. It is shown that if the ''slow'' variation means that the second or higher (crossed or non-crossed) order derivatives of the EPSD function are equal to zero, then the use of STFT (i.e., the approach proposed by Priestley (1965) but without smoothing in time) results in the estimated EPSD function equal to its target, except the residual that depends on the multiplication of two first order derivatives and the adopted window for the transform. In general, the non-zero second- and higher-order derivatives all contribute to the residual. This provides an approach to quantify the effect of the "slow" variation used in the evolutionary process on the estimated EPSD function. The observations that are made based on the use of STFT are applicable if (a generalized) ST is employed. However, if the derived equation for estimating the EPSD function by using CWT is employed, the non-zero first-order derivatives of the EPSD function could contribute to the residual as well. For the considered cases, it is shown that there is a direct relationship between the estimated EPSD function by using a transform and the power distribution function that is represented in the transform domain for the considered transform.

In the following, we first provide the background on the approaches proposed by Priestley (1965) and Spanos and Failla (2004) for estimating the EPSD function. The background serves as the basis for the discussion and further development. We then derive the simple equations for estimating the EPSD function and its corresponding residual. We illustrate its application, especially the estimation of the EPSD function using actual as well as simulated ground motion records, and actual wind speed records.

## 2.0 Background on the estimation of the EPSD function

In this section, we summarize the key aspects of the approaches proposed by Priestley (1965) and Spanos and Failla (2004) to estimate the EPSD function for a given signal. The summary serves as the basis for discussing their differences. It also provides the basis for developing the approach to quantify the effect of the degree of smoothness on the estimated EPSD function.

Consider the evolutionary process $x(t)$ defined as (Priestley 1965),

$$x(t) = \int_{-\infty}^{\infty} A(f,t) e^{i2\pi f t} dZ(f), \tag{1}$$

where $f$ is the frequency, $dZ(f)$ has an orthogonal increment, $E(dZ(f) dZ^*(f')) = 0$ if $f \neq f'$ and $E(|dZ(f)|^2) = S_0(f) df$ if $f = f'$, where $*$ denotes the complex conjugate. Heuristically, one may view the definition as $x(t) = \int_{-\infty}^{\infty} A(f,t) e^{i2\pi f t} e^{i2\pi \theta(f)} df$, where $\theta(f)$ is an $f$-indexed random phase angle and $S_0(f) = 1$, to aid the understanding of the evolutionary process, although strictly speaking this expression is incorrect. The EPSD function of $x(t)$, $S_E(f,t)$, is given by,

$$S_E(f,t) = |A(f,t)|^2 S_0(f). \tag{2}$$

Priestley (1965) considered that $x(t)$ can be decomposed in the time-frequency domain by applying STFT. The STFT coefficient of $x(t)$, $x_{STFT}(f,\tau)$, is given by,

$$x_{STFT}(f,\tau) = STFT(x(t)) = \int_{-\infty}^{\infty} x(t) v(\tau - t) e^{-i2\pi f t} dt = \int_{-\infty}^{\infty} v(s) x(\tau - s) e^{-i2\pi f (\tau - s)} ds, \tag{3}$$

where STFT( ) denotes the STFT operator, $v(t)$ is the window for STFT. Substituting Eq. (1) into Eq. (3), (see equations (6.4) and (6.5) in Priestley (1965) but with different notations), after some algebraic manipulation, one has,

$$x_{STFT}(f,\tau) = \int_{-\infty}^{\infty} v_{A-FT}(\xi,f,\tau)A(\xi+f,\tau)e^{i2\pi\xi\tau}dZ(\xi+f) \tag{4}$$

where $v_{A-FT}(\xi,\xi+f,\tau) = \int_{-\infty}^{\infty}\left[\dfrac{A(\xi+f,\tau-s)}{A(\xi+f,\tau)}\right]v(s)e^{-i2\pi\xi s}ds$. Eq. (4) indicates that $x_{STFT}(f,\tau)$ is the average of the process within a band of frequencies in the region of $f$ and an interval of time in the neighborhood of $\tau$ if the window $v(t)$ is narrow. According to Priestley, if $A(\xi+f,\tau-s)$ is slowly varying compared to $v(s)$ for each $\tau$ and $\xi+f$, there is an important case where, for each $f$ and $\tau$, $v_{A-FT}(\xi,\xi+f,\tau)$ reduces to $\hat{v}(\xi)$, where $\hat{v}(\xi)$ represents the Fourier transform (FT) of $v(t)$ (in the following a symbol with circumflex is used to denote its FT). In other words, this approximation is focused on the selection of a narrow window (as compared to that for $A(f,t)$ along the time-axis) for viewing the evolutionary process. In particular, a box window (i.e., $v(t) = 1/\sqrt{2h}$ for $|t| \leq h$ and equal to zero for $t > h$) was emphasized in Priestley (1965). $v(t)$ is a normalized box window such that $\int_{-\infty}^{\infty}|\hat{v}(\xi)|^2 d\xi = 1$.

Based on the above considerations and taking into account the orthogonality property of $dZ(f)$, one obtains,

$$\begin{aligned}&E\left(x_{STFT}(f,\tau)x_{STFT}^*(f,\tau)\right)\\&\approx \int_{-\infty}^{\infty}|\hat{v}(\xi)|^2|A(\xi+f,\tau)|^2 S_0(\xi+f)d\xi = \int_{-\infty}^{\infty}|\hat{v}(\xi)|^2 S_E(\xi+f,\tau)d\xi\end{aligned} \tag{5}$$

where the approximation comes from using $\hat{v}(\xi)$ for $v_{A-FT}(\xi,\xi+f,\tau)$, and the last equality is based on Eq. (2). From Eq. (5), the consideration that $S_E(f,t)$ is "flat" as compared with $|\hat{v}(\xi)|^2$ results in (see Eq. (9.2) and (9.3) in Priestley (1965) but with different notations),

$$S_E(f,\tau) \approx E\left(x_{STFT}(f,\tau)x_{STFT}^*(f,\tau)\right)/C_n^2 \tag{6}$$

where $C_n^2 = \int_{-\infty}^{\infty}|\hat{v}(\xi)|^2 d\xi$ represents the power normalization constant. Note that this equation is applicable whether a power normalized window is used. The power normalized window defined by $\int_{-\infty}^{\infty}|\hat{v}(\xi)|^2 d\xi = 1$ is considered in Priestley (1965) as mentioned earlier.

For a given digital sample of $x(t)$, $E\left(x_{STFT}(f,\tau)x_{STFT}^*(f,\tau)\right)$ in Eq. (6) may be replaced by $x_{STFT}(f,\tau)x_{STFT}^*(f,\tau)$ and the estimate given on the right-hand side of Eq. (6) is with sampling fluctuation. To reduce the fluctuation, Priestley suggested that a weighting function, $v_T(t)$ is to be applied to calculate the EPSD function in the time-frequency domain, where the sum or integral of the weighting function equal to one is required. The above considerations and the use of $v_T(t)$ and Eq. (6) lead to the estimated $S_E(f,\tau)$ for a given sample that is approximated by $V(f,\tau)$ (see equations (9.2) and (9.7) in Priestley (1965)),

$$V(f,\tau) = \int_{-\infty}^{\infty}\int_{-\infty}^{\infty}|\hat{v}(\xi)|^2|x_{STFT}(\xi+f,t)|^2 v_T(\tau-t)d\xi dt. \tag{7}$$

For further discussion on this approximation in relation to the "width" of $v(t)$, and the width of $A(f,t)$ conditioned on a given $t$ (i.e., the "width" of $A(f,t)$ where $f$ is mapped to time for given $t$), the interested reader is referred to Priestley (1965). It should be noted that, because of the uncertainty principle, the more accurately one tries to determine the EPSD function as a function of time, the less accurately one determines it as a function of frequency, and vice versa. The window in STFT is frequency independent, that is, STFT could not achieve a good time localized resolution at high frequencies and good resolution at low frequencies simultaneously. To overcome this, one could use ST which uses a time and frequency dependent window. The application of ST is to be discussed in the following section.

Rather than using STFT, Spanos and Failla (2004) suggested the use of CWT to estimate the EPSD function. CWT can be written as (Daubechies 1992; Percival and Walden 2000),

$$x_W(s,\tau) = WT(x(t)) = \frac{1}{\sqrt{|s|}} \int_{-\infty}^{\infty} x(t)\psi^*\left(\frac{t-\tau}{s}\right)dt = \sqrt{|s|}\int_{-\infty}^{\infty} \hat{x}(f)\hat{\psi}^*(sf)e^{i2\pi f\tau}df \tag{8}$$

CWT can be inverted if the admissibility condition $0 < C_\psi = \int_{-\infty}^{\infty}\left(\left|\hat{\psi}(f)\right|^2 / |f|\right)df < \infty$ is satisfied.

Substituting Eq. (1) into Eq. (8), considering the time localization properties of the wavelet $\psi\left((t-\tau)/s\right)$ in the vicinity of the time instant $\tau$, and assuming that the rate of change of $A(f,\tau)$ is "slower" with respect to the effective duration of the wavelet $\psi\left((t-\tau)/s\right)$, Spanos and Failla (2004) proposed the following approximation,

$$\begin{aligned} x_W(s,\tau) &= \int_{-\infty}^{\infty} \frac{1}{\sqrt{|s|}}\left[se^{i2\pi\eta\tau}\int_{-\infty}^{\infty} A(\eta,t)\psi^*\left(\frac{t-\tau}{s}\right)e^{i2\pi s\eta\frac{t-\tau}{s}}d\frac{t-\tau}{s}\right]dZ(\eta) \\ &\approx \int_{-\infty}^{\infty} \frac{1}{\sqrt{|s|}}A(\eta,\tau)\left[se^{i2\pi\eta\tau}\int_{-\infty}^{\infty} \psi^*\left(\frac{t-\tau}{s}\right)e^{i2\pi s\eta\frac{t-\tau}{s}}d\frac{t-\tau}{s}\right]dZ(\eta) = \int_{-\infty}^{\infty} A(\eta,\tau)\frac{s}{\sqrt{|s|}}\hat{\psi}^*(s\eta)e^{i2\pi\eta\tau}dZ(\eta) \end{aligned} \tag{9}$$

In writing this equation, FT of the $\psi^*(t)$ equals $(\hat{\psi}(-f))^*$ (which is denoted as $\hat{\psi}^*(-f)$) is used. The approximation shown in Eq. (9) can be accurate if $A(\eta,\tau)$ is "flat" along the time axis. That is, it is expected that the accuracy of the approximation depends on the rate of change of $A(\eta,t)$ (with respect to time) as well as the time localization properties $\psi\left((t-\tau)/s\right)$ in the vicinity of the time instant $\tau$. From Eqs. (2) and (9), one has,

$$E\left[x_W(s,\tau)x_W^*(s,\tau)\right] \approx \int_{-\infty}^{\infty} |A(\eta,\tau)|^2 S_0(\eta)\left|\sqrt{|s|}\hat{\psi}^*(s\eta)\right|^2 d\eta = \int_{-\infty}^{\infty} S_E(\eta,\tau)\left|\sqrt{|s|}\hat{\psi}^*(s\eta)\right|^2 d\eta \tag{10}$$

To estimate $S_E(\eta,t)$ from $E\left[x_W(s,\tau)x_W^*(s,\tau)\right]$, Spanos and Failla (2004) assumed that,

$$S_E(\eta,\tau) = \sum_{j=1}^{N_S} c_j(\tau)\left|\hat{\psi}^*(s_j\eta)\right|^2 \tag{11}$$

which could be viewed as the non-negative matrix factorization of $S_E(\eta,t)$. By substituting Eq. (11) into Eq. (10), a set of equations, each for an assigned $s$ value, is obtained. For each considered time $\tau$, $c_j(\tau)$ for $j=1,...,N_S$, are then evaluated by solving the system of equations established based on Eqs. (10) and (11). Examples of using this approach were presented in Spanos and Failla (2004) and Huang and Chen (2009).

Based on the above summary, it is clear that while the approach given by Priestley (1965) is focused on selecting a frequency-independent window in STFT to decompose the signal and estimate the EPSD function, the approach based on the CWT developed by Spanos and Failla (2004) is focused on projecting the signal in the wavelet domain and estimating the EPSD function. The approximation in both approaches assumed a "slowly varying" or "flat" EPSD function in the vicinity of the time instant τ, where the assessment of the EPSD function is of interest. The quantification of the slow variation is not provided, and the effect of the rate of change of $A(f,\tau)$ on the residual of the estimated EPSD function is not given or discussed. The derivation of the residual and the relation between the estimated EPSD and the power distribution that is represented in the transform domain for the considered transform is addressed in the following sections.

### 3.0 Influence of the smoothness of modulation function on the estimated EPSD function
### 3.1 Considered transforms and Taylor series expansion

To quantify the effect of the time-varying rate of the modulation function on the estimated EPSD function, we assume that the derivatives of $A(\eta,t)$ at an expansion point $(f, \tau)$ in the frequency and time domain exist. The Taylor series expansion of $A(\eta,t)$ at $(f, \tau)$ is expressed as,

$$A(\eta,t) = \sum_{j=0}^{\infty} \sum_{0 \leq l,m \leq j; l+m=j} C_{l,m} (\eta-f)^l (t-\tau)^m, \tag{12}$$

where

$$C_{l,m} = \frac{1}{l!m!} \frac{\partial^{l+m} A(\eta,t)}{\partial^l \eta \partial^m t} \Big|_{\eta=f, t=\tau}. \tag{13}$$

Since the separation of $A(\eta,t)$ and $S_0(\eta)$ is usually unknown for given samples of a process that is assumed to be an evolutionary process, it is considered that the estimation of the EPSD function amount to estimate $A(\eta,t)$ which contains $S_0(\eta)$. In other words, it is considered in the following that $S_0(\eta) = 1$ (i.e., $E(|dZ(\eta)|^2) = d\eta$), and consequently, the EPSD function $S_E(f,\tau)$ equals $|A(f,\tau)|^2$.

For the derivation, we will consider STFT shown in Eq. (3), CWT shown in Eq. (8), and ST (Stockwell et al. 1996), which is defined as,

$$x_S(f,\tau) = ST(x(t)) = \int_{-\infty}^{\infty} x(t) w(f, \tau-t) e^{-i2\pi ft} dt, \tag{14}$$

where $x_S(f,\tau)$ is the ST coefficient, $ST(\ )$ represents the ST operator, and

$$w(f,\tau) = |f| e^{-f^2\tau^2/(2\kappa^2)} / \left(\sqrt{2\pi}\kappa\right), \tag{15}$$

is a window that depends on the time and frequency, and κ is a parameter. ST is also known as generalized ST if κ is not equal to one (Pinnegar and Mansinha 2003). ST can be viewed as a hybrid of STFT and CWT. Also, ST can be further generalized by replacing κ with a positive function of $f$, $K(f)$. Examples of such a generalization can be found in McFadden et al. (1999) and Sejdic et al. (2007).

### 3.2 Estimation based on STFT

By substituting Eq. (1) into Eq. (3) but representing $A(f,t)$ with its Taylor series expansion,

we have,

$$x_{STFT}(f,\tau) = \int_{-\infty}^{\infty}\int_{-\infty}^{\infty}\left[\sum_{j=0}^{\infty}\sum_{0\le l,m\le j, l+m=j} C_{l,m}(\eta-f)^l(t-\tau)^m\right]e^{i2\pi\eta t}v(\tau-t)e^{-i2\pi ft}dt\,dZ(\eta)$$

$$= \int_{-\infty}^{\infty}\left[\sum_{j=0}^{\infty}\sum_{0\le l,m\le j, l+m=j} C_{l,m}(\eta-f)^l \int_{-\infty}^{\infty} s^m v(-s)e^{i2\pi(\eta-f)s}ds\right]e^{i2\pi(\eta-f)\tau}dZ(\eta) \qquad (16)$$

$$= \int_{-\infty}^{\infty}\left[\sum_{j=0}^{\infty}\sum_{0\le l,m\le j, l+m=j} C_{l,m}(\eta-f)^l M_m(\eta-f)\right]e^{i2\pi(\eta-f)\tau}dZ(\eta)$$

where

$$M_m(\eta-f) = \int_{-\infty}^{\infty} s^m v(-s)e^{i2\pi(\eta-f)s}ds. \qquad (17)$$

Note that

$$M_0(\eta-f) = \int_{-\infty}^{\infty} v(-s)e^{i2\pi(\eta-f)s}ds = \hat{v}(\eta-f), \qquad (18)$$

it represents FT of $v(t)$ evaluated at $(\eta-f)$.

Using the obtained STFT coefficients shown in Eq. (16) and the orthogonal property of $dZ(\eta)$, we have,

$$E\left(x_{STFT}(f,\tau)x^*_{STFT}(f,\tau)\right)$$

$$= \int_{-\infty}^{\infty}\left[\sum_{j_1=0}^{\infty}\sum_{0\le l_1,m_1\le j_1, l_1+m_1=j_1} C_{l_1,m_1}(\eta-f)^{l_1}M_{m_1}(\eta-f)\right]\left[\sum_{j_2=0}^{\infty}\sum_{0\le l_2,m_2\le j_2, l_2+m_2=j_2} C_{l_2,m_2}(\eta-f)^{l_2}M_{m_{12}}(\eta-f)\right]^* d\eta$$

$$= \sum_{j_1=0}^{\infty}\sum_{0\le l_1,m_1\le j_1, l_1+m_1=j_1}\sum_{j_2=0}^{\infty}\sum_{0\le l_2,m_2\le j_2, l_2+m_2=j_2} C(l_1,m_1,l_2,m_2)$$

$$= S_E(f,\tau)C_n^2 + R_{STFT}(f,\tau)C_n^2 \qquad (19)$$

where

$$R_{STFT}(f,\tau) = \frac{1}{C_n^2}\sum_{\substack{j_1=0 \\ j_1+j_2>0}}^{\infty}\sum_{0\le l_1,m_1\le j_1, l_1+m_1=j_1}\sum_{j_2=0}^{\infty}\sum_{0\le l_2,m_2\le j_2, l_2+m_2=j_2} C(l_1,m_1,l_2,m_2), \qquad (20)$$

and

$$C(l_1,m_1,l_2,m_2) = C_{l_1,m_1}C^*_{l_2,m_2}\int_{-\infty}^{\infty}\xi^{l_1+l_2}M_{m_1}(\xi)M^*_{m_2}(\xi)d\xi, \qquad (21)$$

The identification of the first term on the right-hand side of Eq. (19) is based on that

$$\int_{-\infty}^{\infty} M_0(\eta-f)M_0^*(\eta-f)d\eta = \int_{-\infty}^{\infty}|\hat{v}(f-\eta)|^2 d\eta = \int_{-\infty}^{\infty}|\hat{v}(\xi)|^2 d\xi = C_n^2, \qquad (22)$$

where the last equality represents the power normalization constant mentioned in Eq. (6), and that $C_{0,0}C_{0,0}^* = |A(f,\tau)|^2$ (see Eqs. (12) and (13), where $S_E(f,\tau) = |A(f,\tau)|^2$ since $S_0(f)=1$ (see explanation to Eq. (13)).

Eq. (19) can be re-written as,

$$S_E(f,\tau) = E\left(x_{STFT}(f,\tau)x^*_{STFT}(f,\tau)\right)/C_n^2 - R_{STFT,f}(f,\tau) = S_{STFT,f}(f,\tau) - R_{STFT}(f,\tau) \qquad (23)$$

where $S_{STFT,f}(f,\tau) = E\left(x_{STFT}(f,\tau)x^*_{STFT}(f,\tau)\right)/C_n^2$ represents the time-frequency dependent PSD function in the STFT domain.   This indicates that the approach proposed by Priestley (1965), which is shown in Eq. (6), only used the first term on the right-hand side of Eq. (23) as the estimate of the EPSD function, which equals the time-frequency dependent PSD function in the STFT domain.   The derived residual corresponding to the estimate was not given in Priestley (1965).

The derivate equation for the residual $R_{STFT}(f,\tau)$ depends on the derivatives of $A(\eta,t)$. In particular, the residual term by including only the first order derivative of $A(\eta,t)$, denoted as $R_{STFT}(f,\tau;1)$, can be written as,

$$R_{STFT}(f,\tau;1)C_n^2 = C_{0,1}C^*_{0,0}\int_{-\infty}^{\infty} M_1(\xi)M^*_0(\xi)d\xi + C_{1,0}C^*_{0,0}\int_{-\infty}^{\infty} \xi M_0(\xi)M^*_0(\xi)d\xi$$
$$+C_{0,0}C^*_{0,1}\int_{-\infty}^{\infty} M_0(\xi)M^*_1(\xi)d\xi + C_{0,0}C^*_{1,0}\int_{-\infty}^{\infty} \xi M_0(\xi)M^*_0(\xi)d\xi \qquad (24)$$

Since FT of an even function is an even function, and FT of an odd function is an odd function but imaginary, $M_0(\xi)$ is an even function and $M_1(\xi)$ is an imaginary odd function if an even window function $v(t)$ such as the box window or Gaussian window is considered.   It can be shown that,

$$R_{STFT}(f,\tau;1) = 0. \qquad (25)$$

For real-valued modulation function $A(\eta,\tau)$ and an even window function $v(t)$, it can be shown that the residual term by including only the second order derivatives (crossed or non-crossed terms, or multiplication of two of the first order derivatives), denoted as $R_{STFT}(f,\tau;2)$, is given by,

$$R_{STFT}(f,\tau;2) = \left(C_{1,0}C_{1,0} + 2C_{2,0}C_{0,0}\right)r(2,2,0,0) + C_{0,1}C_{0,1}r(0,0,2,0) + 2C_{0,2}C_{0,0}r(0,1,0,1) \qquad (26)$$

where

$$r(k,l,m,n) = \frac{1}{C_n^2}\int_{-\infty}^{\infty} \xi^k (M_0(\xi))^l |M_1(\xi)|^m (M_2(\xi))^n d\xi, \qquad (27)$$

Analogously, we can identify and define $R_{STFT}(f,\tau;j)$, for $j = 3,4,\cdots$, based on Eqs. (20) and (21).   Although this exercise is not pursued further, the above-described formulation provides a direct relation between the residual $R_{STFT}(f,\tau)$ in the estimated EPSD function and the smoothness of the modulation function that is defined by its derivatives.   Moreover, although one could include $R_{STFT}(f,\tau;j)$ for $j=1,\cdots,n,$ in estimating the EPSD, and considering $R_{STFT}(f,\tau;j)$ for $j=n,n+1,\cdots,$ as the residual, this is not investigated numerically in the following.

To appreciate the effect of the adopted windows on $R_{STFT}(f,\tau;2)$, numerical evaluation of the ratio $r(k,l,m,n)$ shown in Eq. (27) is carried out by considering the box window defined as $v(t) = 1/(2h)$ for $|t| \leq h$ and equal to zero for $t > h$, and the Gaussian window defined as $v(t) = e^{-\tau^2/(2\sigma^2)}/(\sqrt{2\pi}\sigma)$, where $h$ and $\sigma$ are parameters for the considered windows.

For the calculation, we note that $M_0(\eta - f)$ for the box window is given by,

$$M_0(\eta - f) = \int_{-h}^{h} \frac{1}{2h}e^{i2\pi(\eta - f)s}ds = \frac{\sin(2\pi(\eta - f)h)}{2\pi(\eta - f)h}, \qquad (28)$$

and for the Gaussian window is given by,

$$M_0(\eta-f) = \int_{-\infty}^{\infty} \frac{1}{\sqrt{2\pi}\sigma} e^{-s^2/(2\sigma^2)} e^{i2\pi(\eta-f)s} ds = \exp\left(-2\pi^2\sigma^2(\eta-f)^2\right), \tag{29}$$

These and Eq. (22) result in $C_n^2$ equals $1/(2h)$ for the box window and $1/(2\sigma\sqrt{\pi})$ for the Gaussian window. For easy reference, the approximate estimate of the EPSD function shown in Eq. (23) and the required constant $C_n^2$ for the considered windows are summarized in Table 1.

The calculated values of $r(k,l,m,n)$ are shown in Figures 1 and 2, for the box window and Gaussian window, respectively. Figure 1 indicates that $r(2,2,0,0)$ decreases as $h$ increases. Therefore, if $A(\eta,t)$ is frequency-varying (i.e., not flat in frequency or $C_{1,0}$ and $C_{2,0}$ are not equal to zero), the use of increased window width $h$ decreases the residual contributed by frequency-varying EPSD function. Figure 1 shows that $r(0,0,2,0)$ and $r(0,1,0,1)$ increase as $h$ increases, indicating that the residual is likely to be increased due to time-varying $A(\eta,t)$ (i.e., $C_{0,1}$ and $C_{0,2}$ are not equal to zero) as $h$ increases. This indicates that the selection of $h$ is crucial for reducing the residual in the time-frequency domain to a tolerable level if $A(\eta,t)$ varies in time and frequency. The same conclusions can be made from the results presented in Figure 2, except that in this case, the controlling parameter is $\sigma$ rather than $h$.

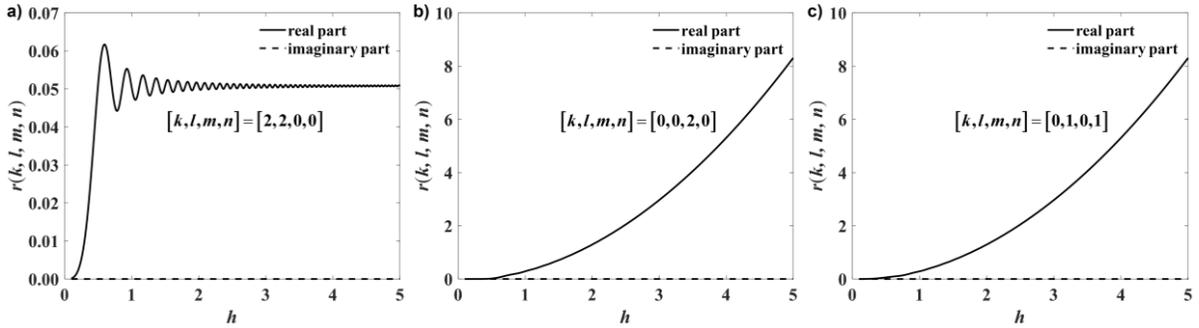

Figure 1. Evaluated ratios, $r(k,l,m,n)$, by considering STFT with the box window.

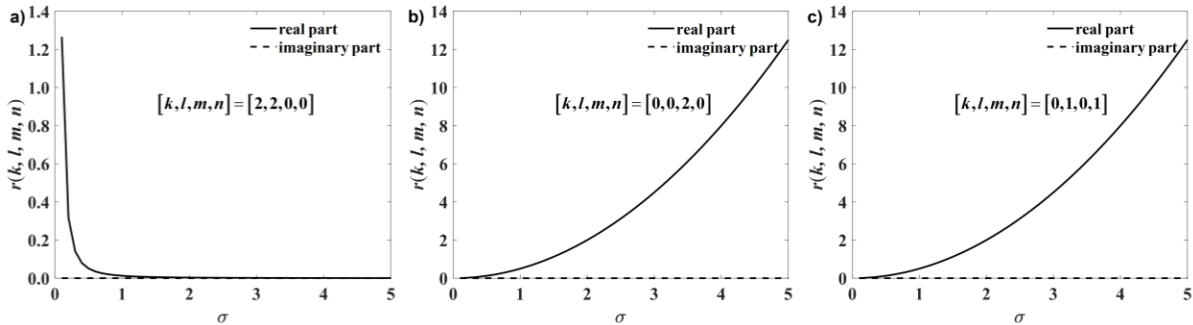

Figure 2. Evaluated ratios, $r(k,l,m,n)$, by considering STFT with the Gaussian window.

Table 1.  Summary of the derived equations for the considered transforms.

| Transform | Window or wavelet | Time-frequency or -scale dependent PSD function in the transform domain. | Power normalization factor | Simple estimate for the EPSD function $S_E(f,\tau)$ | Comments |
|---|---|---|---|---|---|
| STFT | $v(t) = 1/(2h)$ $\|t\| \leq h$ $v(t) = e^{-\tau^2/(2\sigma^2)}/(\sqrt{2\pi}\sigma)$ | $S_{STFT,f}(f,\tau) = \dfrac{E\left(x_{STFT}(f,\tau)x^*_{STFT}(f,\tau)\right)}{C_n^2}$ | $C_n^2 = 1/(2h)$ $C_n^2 = 1/(2\sigma\sqrt{\pi})$ | $S_{STFT,f}(f,\tau)$ Priestley (1965) used the smoothed $S_{STFT,f}(f,\tau)$ as the estimate. | $R_{STFT}(f,\tau;1) = 0$ $R_{STFT}(f,\tau;2)$ is given in Eq. (26). See also Figures 1 and 2. |
| ST | $w(f,\tau) = \dfrac{\|f\|e^{-f^2\tau^2/(2\kappa^2)}}{\sqrt{2\pi}\kappa}$ $w(f,\tau) = \dfrac{\|f\|e^{-f^2\tau^2/(2K^2(f))}}{\sqrt{2\pi}K(f)}$ | $S_{Sf}(f,\tau) = \dfrac{E\left(x_S(f,\tau)x^*_S(f,\tau)\right)}{\|f\|D_{K(f)}}$ The first case is a special case of the second case, where $K(f) = \kappa$. | $\|f\|D_{K(f)} = \|f\| \times \displaystyle\int_{-\infty}^{\infty} e^{-(2\pi K(f)(\zeta-1))^2}\dfrac{d\zeta}{\|\zeta\|}$, $D_{K(f)} \approx \dfrac{1}{K(f)\sqrt{4\pi}}$ if $K(f) > 1$. | $S_{Sf}(f,\tau)\dfrac{D_{K(f)}}{C_{nS0}}$ See Eq. (37) for the evaluation of $D_{K(f)}$ and Figure 3. | $R_S(f,\tau;1) = 0$ $R_S(f,\tau;2)$ is given in Eq. (38). See also Figure 4. |
| CWT | HW, $\psi_H(t) = \dfrac{(e^{i2\pi nt} - e^{i2\pi mt})}{(i2\pi t\sqrt{n-m})}$ GMW, $\psi_{0,\beta,\gamma}(t)$ is defined through $\hat{\psi}_{0,\beta,\gamma}(f) = U(f) \times a_{\beta,\gamma}(2\pi f)^\beta e^{-(2\pi f)^\gamma}$ | $S_{Ws}(s,\tau) = \dfrac{E(x_W(s,\tau)x^*_W(s,\tau))}{C_W^2}$, $S_{Wf}(f,\tau) = S_{Ws}(s,\tau)\left\|\dfrac{ds}{df}\right\|$ where $f = g(s) = f_0/s$ and $s = f_0/f$. $f_0 = (n+m)/2$ for HW and $f_0 = (\beta/\gamma)^{1/\gamma}/(2\pi)$ for GMW. | $C_W^2 = \|s\sqrt{C_\psi}\|^2$, $C_\psi = \dfrac{\ln(n/m)}{(n-m)}$, for HW, and $C_\psi = \dfrac{2a_{2\beta,\gamma}\Gamma(2\beta/\gamma)}{\gamma}$ for GMW. | $S_{Wf}(f,\tau)\left[\dfrac{C_W^2 f_0}{C_{nW}^2 s^2}\right]$ (see Eq. (50)), $C_W^2 f_0 / (C_{nW}^2 s^2)$ $= \dfrac{\ln(n/m)}{n-m}\dfrac{n+m}{2}$ for HW, and $= \dfrac{(2\beta/\gamma)^{1/\gamma}\Gamma(2\beta/\gamma)}{\Gamma((2\beta+1)/\gamma)}$ for GMW. | For $R_W(f,\tau;1)$ and $R_W(f,\tau;2)$, see Eqs. (51) to (53). $C_{nW}^2 = 1$ for HW, and $C_{nW}^2 = \dfrac{2a_{2\beta,\gamma}\Gamma\left(\dfrac{2\beta+1}{\gamma}\right)}{(2\pi)2^{1/\gamma}\gamma}$ for GMW. For some selected wavelet parameters. $C_W^2 f_0/(C_{nW}^2 s^2) \approx 1$ |

### 3.3 Estimation based on ST

By considering ST and following the same algebraic procedure employed in deriving Eqs. (16) to (21), it can be shown that,

$$x_S(f,\tau) = ST(x(t)) = \int_{-\infty}^{\infty} \left[ \sum_{j=0}^{\infty} \sum_{0 \leq l, m \leq j, l+m=j} C_{l,m}(\eta-f)^l M_{S,m}(\eta-f;f) \right] e^{i2\pi(\eta-f)\tau} dZ(\eta), \quad (30)$$

where

$$M_{S,m}(\eta-f;f) = \int_{-\infty}^{\infty} (s)^m w(f,-s) e^{i2\pi(\eta-f)s} ds, \quad (31)$$

Using the obtained ST coefficients shown in Eq. (27) and the orthogonal property of $dZ(f)$, we have,

$$E\left(x_S(f,\tau) x_S^*(f,\tau)\right)$$

$$= \int_{-\infty}^{\infty} \left[ \sum_{j_1=0}^{\infty} \sum_{0 \leq l_1, m_1 \leq j_1, l_1+m_1=j_1} C_{l_1,m_1} \xi^{l_1} M_{S,m_1}(\xi) \right] \left[ \sum_{j_2=0}^{\infty} \sum_{0 \leq l_2, m_2 \leq j_2, l_2+m_2=j_2} C_{l_2,m_2} \xi^{l_2} M_{S,m_{12}}(\xi) \right]^* d\xi, \quad (32)$$

$$= \sum_{j_1=0}^{\infty} \sum_{0 \leq l_1, m_1 \leq j_1, l_1+m_1=j_1} \sum_{j_2=0}^{\infty} \sum_{0 \leq l_2, m_2 \leq j_2, l_2+m_2=j_2} C_S(l_1,m_1,l_2,m_2) = S_E(f,\tau) C_{nS}^2 + R_S(f,\tau) C_{nS}^2$$

where

$$R_S(f,\tau) = \frac{1}{C_{nS}^2} \sum_{\substack{j_1=0 \\ j_1+j_2>0}}^{\infty} \sum_{0 \leq l_1, m_1 \leq j_1, l_1+m_1=j_1} \sum_{j_2=0}^{\infty} \sum_{0 \leq l_2, m_2 \leq j_2, l_2+m_2=j_2} C_S(l_1,m_1,l_2,m_2), \quad (33)$$

$$C_S(l_1,m_1,l_2,m_2) = C_{l_1,m_1} C_{l_2,m_2}^* \int_{-\infty}^{\infty} (\eta-f)^{l_1+l_2} M_{S,m_1}(\eta-f;f) M_{S,m_2}^*(\eta-f;f) d(\eta-f)$$

$$= C_{l_1,m_1} C_{l_2,m_2}^* \int_{-\infty}^{\infty} (\xi)^{l_1+l_2} M_{S,m_1}(\xi;f) M_{S,m_2}^*(\xi;f) d\xi \quad (34)$$

and,

$$C_{nS}^2 = \int_{-\infty}^{\infty} M_{S,0}(\eta-f;f) M_{S,0}(\eta-f,f) d\eta = |f| C_{nS0}, \quad (35)$$

where $C_{nS0} = 1/(\kappa\sqrt{4\pi})$.

Note that it was shown (Hong 2021) that the energy-preserving time-frequency dependent PSD function in the ST domain can be defined as,

$$S_{Sf}(f,\tau) = \frac{E\left(x_S(f,\tau) x_S^*(f,\tau)\right)}{\sqrt{|f| D_\kappa} \sqrt{|f| D_\kappa}}, \quad (36)$$

where $D_\kappa = \int_{-\infty}^{\infty} \exp\left(-(2\pi\kappa(\zeta-1))^2\right) \frac{d\zeta}{|\zeta|}$, which is approximately equal to $1/(\kappa\sqrt{4\pi})$ (i.e., $C_{nS0}$) with an error less than about 1.5% if $\kappa$ is greater than one. However, the difference between $D_\kappa$ and $C_{nS0}$ increases drastically as $\kappa$ decreases and tends to zero (i.e., the effective window width decreases). This is shown in Figure 3.

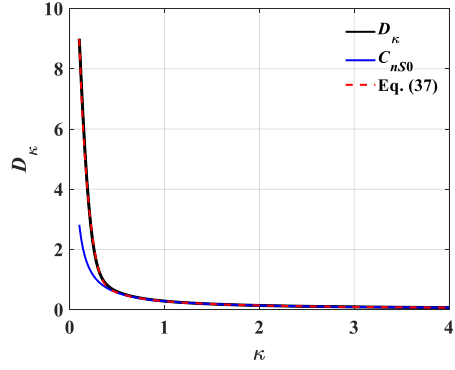

Figure 3.  Comparison of $D_\kappa$ and its approximation.  Note that the horizontal and vertical axes can also be viewed as $D_{\mathrm{K}(f)}$ and K($f$), respectively.

A simple regression analysis indicates that $D_\kappa$ could be better approximated by,

$$D_\kappa \approx (1+2.3e^{-70\kappa^{3.14}})C_{nS0}. \tag{37}$$

for $\kappa$ greater than 0.1. The adequacy of this approximation is shown in Figure 3 as well.

Based on the above, Eq. (32) can be re-written as,

$$S_E(f,\tau) = E\!\left(x_S(f,\tau)x_S^*(f,\tau)\right)/C_{nS}^2 - R_S(f,\tau) = S_{Sf}(f,\tau)\frac{D_\kappa}{C_{nS0}} - R_S(f,\tau). \tag{38}$$

It indicates that if the time-frequency dependent PSD function $E\!\left(x_S(f,\tau)x_S^*(f,\tau)\right)/C_{nS}^2$ is used to estimate $S_E(f,\tau)$, or $S_{Sf}(f,\tau)D_\kappa/C_{nS0}$ the residual term $R_S(f,\tau)$ is associated with the derivatives of $A(\eta,t)$.  Eq. (38) also indicates that the EPSD function can be approximated by the time-frequency dependent $S_{Sf}(f,\tau)$ that is obtained directly from the ST analysis and a scaling factor $D_\kappa/C_{nS0}$, which is practically equal to 1 for $\kappa$ greater than about one.

Similar to the case of using STFT, Eq. (38) shows that the residual term by including only the first order derivative of $A(\eta,t)$, denoted as $R_S(f,\tau;1)$, equals zero.  Also, the residual term by including only the second order derivatives (crossed or non-crossed terms or multiplication of two of the first order derivative), denoted as $R_S(f,\tau;2)$, for the real-valued modulation function is given by,

$$R_S(f,\tau;2) = (C_{1,0}C_{1,0}+2C_{2,0}C_{0,0})r_S(2,2,0,0) + C_{0,1}C_{0,1}^* r_S(0,0,2,0) + 2C_{0,0}C_{0,2}r_S(0,1,0,1) \tag{39}$$

where

$$r_S(k,l,m,n) = \frac{1}{C_{nS}^2}\int_{-\infty}^{\infty}\xi^k(M_{S,0}(\xi))^l \left|M_{S,1}(\xi)\right|^m (M_{S,2}(\xi))^n d\xi,. \tag{40}$$

Again, the exercise of defining and identifying $R_S(f,\tau;j)$, for $j=3,4,\cdots$, based on Eqs. (33) and (34) is not pursued.  The formulation given in this section so far relates $R_{ST}(f,\tau)$ in the estimated EPSD function and the smoothness of the modulation function that is defined by its derivatives.

To appreciate the effect of the window used in ST on $R_{ST}(f,\tau;2)$, the evaluation of the ratio $r_S(k,l,m,n;f)$ is carried out numerically.  Figure 4 shows that $r_S(2,2,0,0)$ increases as $\kappa$ decreases or $f$ increases. This indicates that a better estimate can be obtained at the low frequencies by increasing $\kappa$. The figure also shows that $r_S(0,0,2,0)$ and $r_S(0,1,0,1)$ decrease as $\kappa$ decreases or $f$ increases. Therefore, a reduced residual at the high frequencies is likely to result if $\kappa$ is decreased.

It can be easily shown that the above formulation is equally applicable if the window parameter for ST, $\kappa$, used in Eq. (15) is replaced by the positive function $K(f)$ (i.e., using a generalized ST). In such a case, the new window needs to be employed in Eq. (31) to evaluate $M_{S,m}(\eta-f;f)$, and the required $C_{nS0}$ in Eq. (36) becomes $C_{nS0}=1/(K(f)\sqrt{4\pi})$. The factor $D_\kappa/C_{nS0}$ can be read from Figure 3, where the horizontal axis, in this case, represents $K(f)$ instead of $\kappa$.

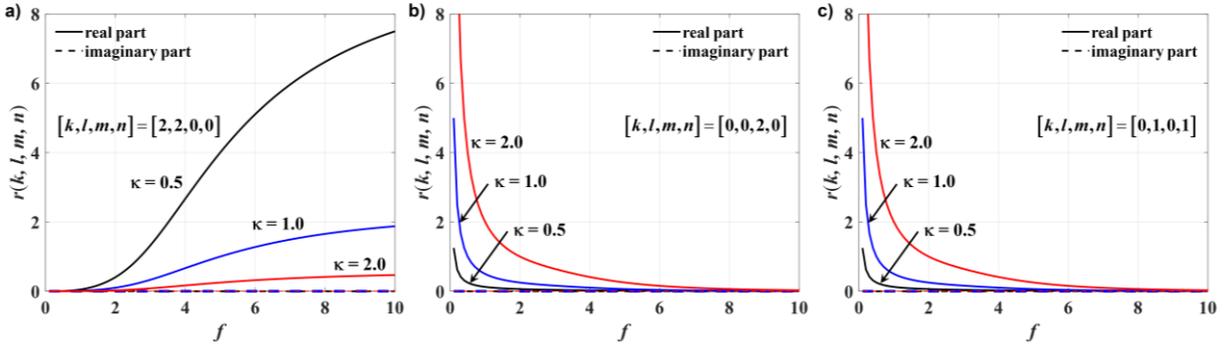

Figure 4. Evaluated ratios, $r_S(k,l,m,n)$, by considering the ST transform (solid line for the real part and dashed line for the imaginary part).

This and the observations based on results shown in Figure 4 suggest that one could select a frequency-dependent $K(f)$, with $K(f)$ greater than 1.0 at the low frequencies and $K(f)$ smaller than 1.0 at the high frequencies to reduce the residual associated with the estimated EPSD function.

### 3.4 Estimation based on CWT

In this section, we consider the use of continuous wavelet transforms in estimating the EPSD function. Again, we follow the procedure described in the previous sections for STFT and ST. By substituting Eq. (1) into Eq. (8) and considering that $A(\eta,t)$ can be expressed in the Taylor series expansion as shown in Eq. (12). This results in,

$$\begin{aligned} x_\mathcal{W}(s,\tau) &= \int_{-\infty}^{\infty}\int_{-\infty}^{\infty} \frac{1}{\sqrt{|s|}} \left[ \sum_{j=0}^{n} \sum_{0\le l,m\le j, l+m=j} C_{l,m}(\eta-f)^l (t-\tau)^m \right] e^{i2\pi\eta t} \psi^*\left(\frac{t-\tau}{s}\right) dt\, dZ(\eta) \\ &= \int_{-\infty}^{\infty} \left[ \sum_{j=0}^{n} \sum_{0\le l,m\le j, l+m=j} C_{l,m}(\eta-f)^l \frac{s^m}{\sqrt{|s|}} \int_{-\infty}^{\infty} \left(\frac{u}{s}\right)^m \psi^*\left(\frac{u}{s}\right) e^{-i2\pi(-\eta)u} du \right] e^{i2\pi\eta\tau} dZ(\eta) \qquad (41)\\ &= \int_{-\infty}^{\infty} \left[ \sum_{j=0}^{n} \sum_{0\le l,m\le j, l+m=j} C_{l,m}(\eta-f)^l M_{\mathcal{W},m}(s\eta,s) \right] e^{i2\pi\eta\tau} dZ(\eta) \end{aligned}$$

where $M_{w,0}(s\eta, s) = \left(s/\sqrt{|s|}\right)\hat{\psi}^*(s\eta)$ and

$$M_{w,m}(s\eta, s) = \left(s^{m+1}/\sqrt{|s|}\right)\hat{\psi}^*(s\eta) = s^m M_{w,0}(s\eta, s). \tag{42}$$

At this point, it is instructive to indicate the relation of $s$ and the frequency value $f$ where the Taylor series expansion is taking place for deriving Eq. (41). Firstly, the relation between the frequency and scale that could be selected based on different criteria can differ for some of the wavelet transforms (Lilly and Olhede 2009, 2012). Therefore, for a selected criterion, it is considered that the functions $f = g(s)$ and $s = g^{-1}(f)$ can be used to map the scale to frequency, and the frequency to scale, respectively.

By considering the orthogonal property of $dZ(f)$ and using Eqs. (41) and (42), we have,

$$\frac{E\left(x_w(s,\tau)x_w^*(s,\tau)\right)}{C_w^2}|ds| = \frac{E\left(x_w(s,\tau)x_w^*(s,\tau)\right)}{C_w^2}\left|\frac{ds}{df}\right||df| =$$

$$= \left[\sum_{j_1=0}^{\infty}\sum_{0\leq l_1,m_1\leq j_1, l_1+m_1=j_1}\sum_{j_2=0}^{\infty}\sum_{0\leq l_2,m_2\leq j_2, l_2+m_2=j_2} C_w(l_1,m_1,l_2,m_2)\right] \times \left[\frac{1}{C_w^2}\left|\frac{ds}{df}\right|\right]|df| \tag{43}$$

where $C_w^2 = s^2 C_\psi$, $C_\psi$ is already defined earlier (see Eq. (8) and the condition for its inverse), and,

$$C_w(l_1,m_1,l_2,m_2) = C_{l_1,m_1}C_{l_2,m_2}^* \int_{-\infty}^{\infty} (\eta - f)^{l_1+l_2} M_{w,m_1}(s\eta,s)M_{w,m_2}^*(s\eta,s)d\eta. \tag{44}$$

The first equality in Eq. (43) maps the power distribution in the time-scale domain of the considered CWT to the time-frequency domain, and the second equality services to express the power distribution by taking into account the Taylor series expansion of the EPSD function considered in deriving Eq. (41). In particular, $C_w(0,0,0,0)$ can be written as,

$$C_w(0,0,0,0) = |A(f,\tau)|^2 \int_{-\infty}^{\infty} M_{w,0}(s\eta,s)M_{w,0}^*(s\eta,s)d\eta$$

$$= |A(f,\tau)|^2 \left(s/\sqrt{|s|}\right)^2 \int_{-\infty}^{\infty} |\hat{\psi}^*(s\eta)|^2 d\eta = S_E(f,\tau)C_{nw}^2 \tag{45}$$

where $C_{nw}^2 = \left(s/\sqrt{|s|}\right)^2 \int_{-\infty}^{\infty} |\hat{\psi}^*(s\eta)|^2 d\eta = \int_{-\infty}^{\infty} |\hat{\psi}^*(\eta)|^2 d\eta$, which represents a power normalization constant. Therefore, by noting that $S_E(f,\tau) = |A(f,\tau)|^2$, Eq. (43) can be re-written as,

$$S_E(f,\tau)|df| = \frac{E\left(x_w(s,\tau)x_w^*(s,\tau)\right)}{C_w^2}\left[\frac{C_w^2}{C_{nw}^2}\left|\frac{df}{ds}\right|\right]|ds| - R_w(f,\tau)|df|, \tag{46}$$

where

$$R_w(f,\tau) = \frac{1}{C_{nw}^2} \sum_{\substack{j_1=0 \\ j_1+j_2>0}}^{\infty} \sum_{0\leq l_1,m_1\leq j_1, l_1+m_1=j_1} \sum_{j_2=0}^{\infty} \sum_{0\leq l_2,m_2\leq j_2, l_2+m_2=j_2} C_w(l_1,m_1,l_2,m_2). \tag{47}$$

and $C_w^2 = |s\sqrt{C_\psi}|^2$, in which $C_\psi$ is already defined earlier (see Eq. (8) and the condition for its inverse).

To clarify the meaning of Eq. (46), we note that the power distribution in the time-scale domain of the considered CWT for a given signal (i.e., scalogram), denoted as $S_{w_x}(s,\tau)$, that satisfied the energy preservation criterion equal to $x_w(s,\tau)x_w^*(s,\tau)/C_w^2$ (Daubechies 1992). Therefore, the power distribution for a stochastic process in the time-scale domain of the considered CWT can be defined as,

$$S_{w_s}(s,\tau) = E(x_w(s,\tau)x_w^*(s,\tau))/C_w^2. \tag{48}$$

and its corresponding power distribution in the time-frequency domain for the considered CWT, denoted as $S_{w_f}(f,\tau)$, can then be defined as,

$$S_{w_f}(f,\tau) = \frac{E(x_w(s,\tau)x_w^*(s,\tau))}{C_w^2}\left|\frac{ds}{df}\right|. \tag{49}$$

Consequently, Eq. (46) can be re-written as,

$$S_E(f,\tau) = S_{w_f}(f,\tau)\left[\frac{C_w^2}{C_{nw}^2}\left|\frac{df}{ds}\right|\right] - R_w(f,\tau). \tag{50}$$

This indicates that the EPSD function can be approximated by using the time-frequency dependent PSD function $S_{w_f}(f,\tau)$ and a scaling factor $(C_w^2/C_{nw}^2)/|df/ds|$ (i.e., the first term in Eq. (50)). Note that the first term on the right-hand side can be rewritten simply as $E(x_w(s,\tau)x_w^*(s,\tau))/C_{nw}^2$ with $s = g^{-1}(f)$. That is, the EPSD can be approximated by $E(x_w(g^{-1}(f),\tau)x_w^*(g^{-1}(f),\tau))/C_{nw}^2$. However, the time-frequency interpretation may be less apparent than as shown in Eq. (50). The residual associated with the estimate equals $R_w(f,\tau)$. It is noteworthy that in deriving the first term on the right-hand side of Eq. (50) (i.e., the approximate estimate of the EPSD function), the first term of the Taylor series expansion of $A(\eta,t)$ at the point $(f, \tau)$, where the PSD is to be estimated, is employed. The approach proposed by Spanos and Failla (2004) approximates the integral involving $A(\eta,t)$ over the time domain at $\tau$ (see Eqs. (9) and (10)), and then approximates or assumes that the EPSD function can be represented by the sum of the multiplication of the function of time and function of frequency (see Eq. (11)). However, they did not derive an equation for evaluating the residual corresponding to their proposed estimate of the EPSD function.

Based on our derived equation for the residual shown in Eq. (47), it can be shown that the residual term by including only the first order derivative of the real-valued $A(\eta,t)$, denoted as $R_w(f,\tau;1)$, can be written as,

$$R_w(f,\tau;1) = 2C_{0,0}C_{0,1}r_w(0,1) + 2C_{0,0}C_{1,0}r_w(1,0), \tag{51}$$

where

$$r_w(j,k) = \frac{1}{C_{nw}^2}\int_{-\infty}^{\infty}(\eta-f)^j s^k |M_{w,0}(s\eta,s)|^2 d\eta = \frac{s^k}{C_{nw}^2}\int_{-\infty}^{\infty}(\eta-f)^j \frac{s^2}{|s|}|\hat{\psi}^*(s\eta)|^2 d\eta. \tag{52}$$

This indicates that $r_w(j,k)/s^k$ represents the $j$-th moment of the power distribution of the wavelet with respect to the selected frequency $f$ which depends on $s$.

Since $r_w(1,0)$ and $r_w(0,1)$ may not necessarily be equal to zero, Eq. (51) indicates that unlike the use of STFT and ST, where $R_{STFT}(f,\tau;1)$ and $R_S(f,\tau;1)$ are equal to zero, $R_{\mathcal{W}}(f,\tau;1)$ is not necessarily equal to zero. The residual term by including only the second order derivatives (crossed or non-crossed or multiplication of two of the first order derivative), denoted as $R_{\mathcal{W}}(f,\tau;2)$, for the real-valued $A(\eta,\tau)$ is given by,

$$R_{\mathcal{W}}(f,\tau;2) = (C_{1,0}C_{1,0} + 2C_{0,0}C_{2,0})r_{\mathcal{W}}(2,0) + (2C_{1,0}C_{0,1} + 2C_{0,0}C_{1,1})r_{\mathcal{W}}(1,1) + (C_{0,1}C_{0,1}^* + 2C_{0,0}C_{0,2})r_{\mathcal{W}}(0,2) \tag{53}$$

Eqs. (51) to (53) show that the residual for using Eq. (50) to estimate the EPSD function depends not only on the derivatives of the EPSD function but also on the characteristics of the adopted wavelet transform.

Two wavelets, namely the harmonic wavelet (HW) and generalized Morse wavelet (GMW), are considered to evaluate $r_{\mathcal{W}}(j,k)$ in the following. The CWT with the Morlet wavelet is not considered since its use is equivalent to the use of ST (Ventosa et al. 2008). HW, $\psi_H(t)$, is defined as (Newland (2012)),

$$\psi_H(t) = (e^{i2\pi nt} - e^{i2\pi mt})/(i2\pi t\sqrt{n-m}), \tag{54}$$

and its corresponding FT $\hat{\psi}_H(f)$ is,

$$\hat{\psi}_H(f) = \begin{cases} 1/\sqrt{n-m}, & m \leq f < n \\ 0, & \text{otherwise} \end{cases}, \tag{55}$$

where $m$ and $n$ are real and positive values and $m < n$. For the harmonic wavelet, $C_\psi = \ln(n/m)/(n-m)$ and $C_{n\mathcal{W}}^2 = 1$. By considering $f_0 = (n+m)/2$ and $f = g(s) = f_0/s$, one has,

$$\frac{C_{\mathcal{W}}^2}{C_{n\mathcal{W}}^2}\left|\frac{df}{ds}\right| = \frac{\ln(n/m)}{n-m}\frac{n+m}{2}. \tag{56}$$

For example, if $(m,n)$ equals $(1,2)$, $(C_{\mathcal{W}}^2/C_{n\mathcal{W}}^2)/|df/ds|$ equals 1.040, indicating that $S_{\mathcal{W}f}(f,\tau)$ may be directly used as an approximate estimate of the EPSD function. As $n$ tends to $m$, the right-hand side of Eq. (56) tends to unity.

GMW, $\psi_{0,\beta,\gamma}(t)$, is defined in the frequency domain as (Bayram and Baraniuk 1996; Olhede and Walden 2002; Lilly and Olhede 2012),

$$\hat{\psi}_{0,\beta,\gamma}(f) = U(f)a_{\beta,\gamma}(2\pi f)^\beta e^{-(2\pi f)^\gamma}, \tag{57}$$

where $U(f)$ is the Heaviside function, $a_{\beta,\gamma} = 2(e\gamma/\beta)^{\beta/\gamma}$, $e$ is Euler's number, $\beta$ and $\gamma$ are the two model parameters of GMW. For GMW, $C_\psi = 2a_{2\beta,\gamma}\Gamma(2\beta/\gamma)/\gamma$ and $C_{1\psi} = a_{\beta,\gamma}\Gamma(\beta/\gamma)/\gamma$. $\hat{\psi}_{0,\beta,\gamma}(f)$ attains its maximum value at $(\beta/\gamma)^{1/\gamma}/(2\pi)$ (Hz). By considering $f = g(s) = f_0/s$ with $f_0 = (\beta/\gamma)^{1/\gamma}/(2\pi)$, we have,

$$\frac{C_{\mathcal{W}}^2}{C_{n\mathcal{W}}^2}\left|\frac{df}{ds}\right| = |s\sqrt{C_\psi}|^2\frac{f_0}{s^2}\frac{1}{C_{n\mathcal{W}}^2} = C_\psi f_0 \frac{1}{C_{n\mathcal{W}}^2} = \frac{(2\beta/\gamma)^{1/\gamma}\Gamma(2\beta/\gamma)}{\Gamma((2\beta+1)/\gamma)}, \tag{58}$$

where $C_{n\mathcal{W}}^2 = \int_{-\infty}^{\infty}|\hat{\psi}^*(\eta)|^2 d\eta = 2a_{2\beta,\gamma}\Gamma((2\beta+1)/\gamma)/((2\pi)2^{1/\gamma}\gamma)$ is employed in the above derivation. For example, if $\beta = 20$ and $\gamma = 3$, $(C_{\mathcal{W}}^2/C_{n\mathcal{W}}^2)/|df/ds|$ equals 1.008, indicating that

$S_{wf}(f,\tau)$ could be used directly as an approximate estimate of the EPSD function.

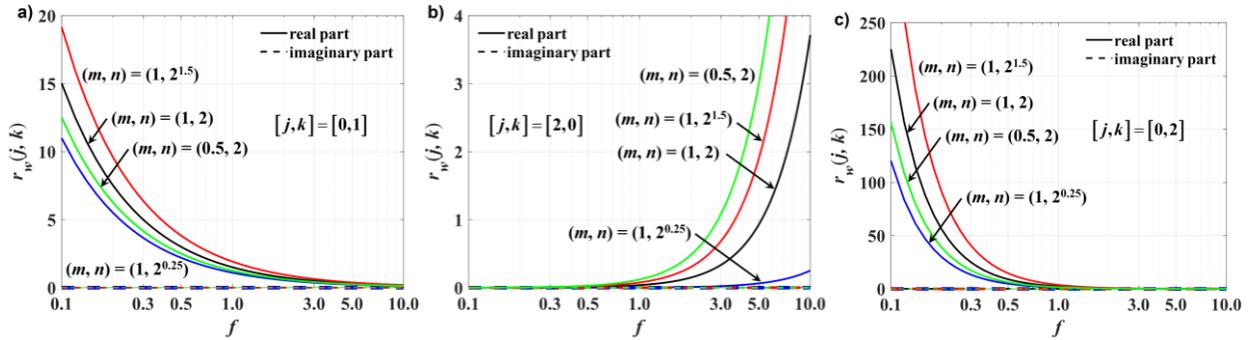

Figure 5. Evaluated $r_w(j,k)$ for $(j, k) = (0,1)$, $(2,0)$, and $(0,2)$ by considering the continuous harmonic wavelet (solid line for the real part and dashed line for the imaginary part).

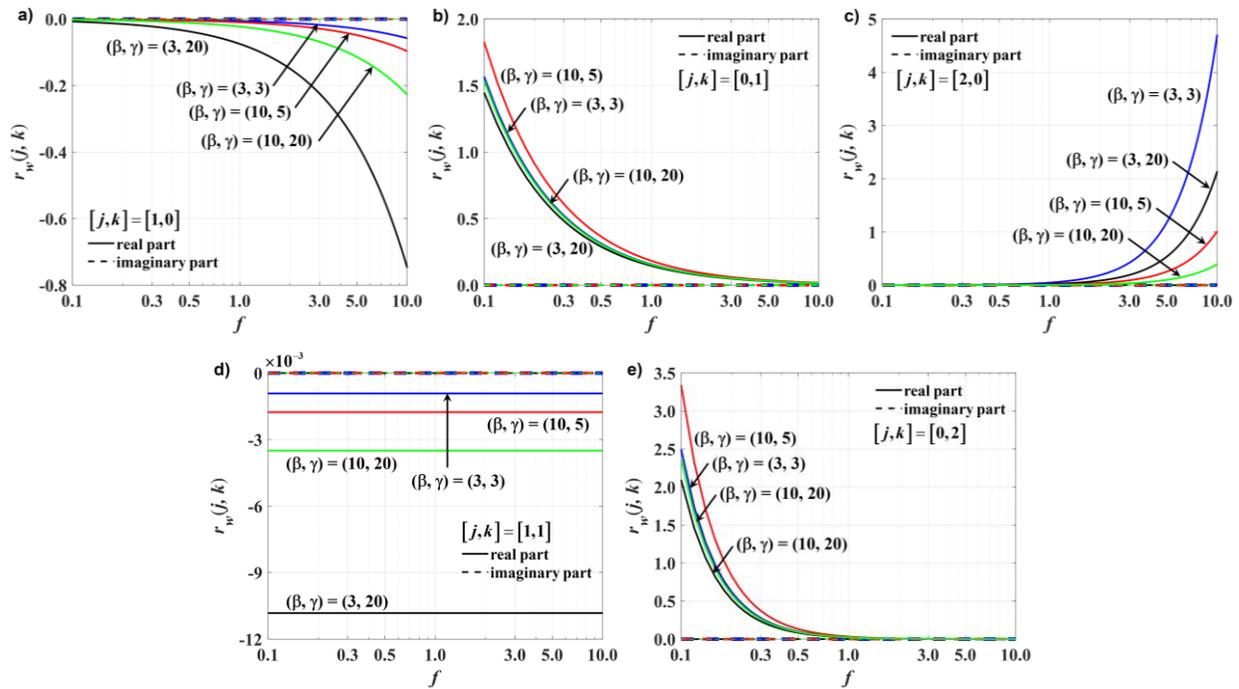

Figure 6. Evaluated $r_w(j,k)$ for $(j, k) = (1,0), (0,1), (2,0), (1,1), (0,2)$ by considering the GMW (solid line for the real part and dashed line for the imaginary part).

The evaluated $r_w(j,k)$ are shown in Figures 5 and 6 for HW and GMW, respectively. For the evaluation, the $(j,k)$ equal to $(0,1)$, $(1,0)$, $(2,0)$, $(1,1)$, $(0,2)$ are considered. For HW, since $r_w(1,0)$ and $r_w(1,1)$ are equal to zero they are not plotted. The values of $r_w(0,1)$ and $r_w(0,2)$ decrease as $f$ increases, indicating that a better estimate of the EPSD function can be obtained if the components of the stochastic process at the low frequencies are stationary. Figure 5 shows that $r_w(2,0)$ increases as $f$ increases, indicating that the accuracy of the estimated EPSD is affected by the trends and curvature of the actual EPSD of the stochastic process with respect to frequency.

## 4.0 Numerical examples illustrating the estimated EPSD function
### 4.1 Comparison of the estimated EPSD from simulated records and its corresponding target

For the numerical example carried out in this section, we consider a set of 10000 synthetic records. For the simulation of the synthetic records, we assume that the seismic ground motions at a shallow alluvial site for an event with a moment magnitude of seven and epicentral distance of 50 km could be modelled as an evolutionary process with (two-sided) EPSD function, we adopt the following PSD function (Cui and Hong 2021a) as the EPSD function,

$$S_E(f,t) = \frac{1}{2} \frac{E_T \lambda_0(t)}{f\sqrt{2\pi}\eta(t)} \exp\left[-\frac{1}{2}\left(\frac{\ln f - \ln F_c(t) + \eta^2(t)/2}{\eta(t)}\right)^2\right], \quad (59)$$

where $t \in [0,T]$, the duration $T$ equals 21.50 s, the total power $E_T$ equals 1478 cm²/s³, $\eta(t) = 0.71$, $F_c(t) = \exp(1.942 - 0.35\ln t)$ (Hz), and

$$\lambda_0(t) = \frac{1}{0.42t\sqrt{1.42\pi}} \exp\left(-\frac{(\ln t - 2.15)^2}{0.18}\right). \quad (60)$$

The value of $S_E(f,t)$ is shown in Figure 7a. By assuming that the ground motion record is a Gaussian process, we simulate 10000 samples of the ground motions using SRM with a sampling frequency of 50 Hz. A typical simulated record is shown in Figure 7b. We apply each of the considered transforms summarized in Table 1 (see the column indicating the simple estimate for the EPSD function. The case of using ST with K(*f*) is not considered to estimate the EPSD function since many options for K(*f*) can be considered.

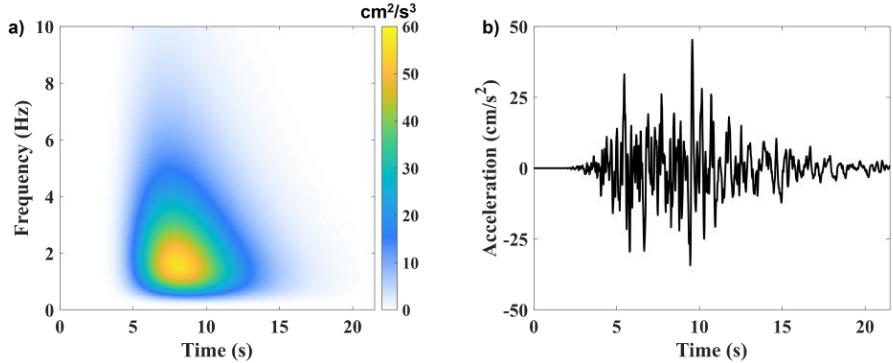

Figure 7. Target EPSD function for the seismic ground motions and a typical sampled ground motion record by using SRM.

By considering each transform listed in Table 1, the decomposition of each of the simulated records is carried out. For the decomposition, the fat Fourier transform is employed for evaluating the FT of the simulated record $\{x(j\Delta_t)\}_{j=0}^{N-1}$. The numerical analysis for using the CWT shown in Eq. (8) is carried out using,

$$x_\mathscr{W}(s_j, q\Delta_t) = \sqrt{|s_j|} \frac{1}{N\Delta_f} \sum_{k=0}^{N-1} \hat{x}(k\Delta_f)\hat{\psi}^*(s_j k\Delta_f) e^{i\frac{2\pi}{N}kq}, \text{ for } j = 0,\cdots,K, \text{ and } q = 0,\cdots,N-1, \quad (61)$$

where $\Delta_f = 1/(N\Delta_t)$; $K+1$ is the total number of scales; $s_j = c_0 s_0^j$ and $j$ represents the $j$-th level of the scale, and $c_0$ and $s_0$ are selected positive parameters for the numerical evaluation.

For each considered transform, the mean and the standard deviation of the calculated EPSD function are shown in Figure 8. As can be observed that the trends of the estimated mean of the PSD function in the time-frequency domain by using each of the considered transforms are very consistent with that of the target EPSD function shown in Figure 7a. The standard deviation of the estimate of the EPSD function reflects the sample-to-sample variability of the simulated record by using SRM. It may also include inaccuracy in the estimation of the EPSD function due to the residual mentioned in the previous sections. The magnitude of the standard deviation of the estimate of the EPSD function from the simulated records is relatively consistent by considering different transforms.

The quantified differences between the estimated mean of EPSD and its corresponding target for each considered transform are also shown in Figure 8 as well. As can be observed from the plotted results, the differences are small, especially for ST and CWT with the GMW. Additional numerical analysis carried out by varying the parameters for the windows and the wavelets indicates that the accuracy in the matching depends on the selected parameters.

This numerical example indicates that the use of each of the selected transforms can result in a reasonable estimate of the EPSD, provided a large number of samples is available. Note that theoretical development of the probability distribution and confidence interval of the estimated EPSD based on a limited number of sampled records for a nonstationary stochastic process is warranted but beyond our intended scope.

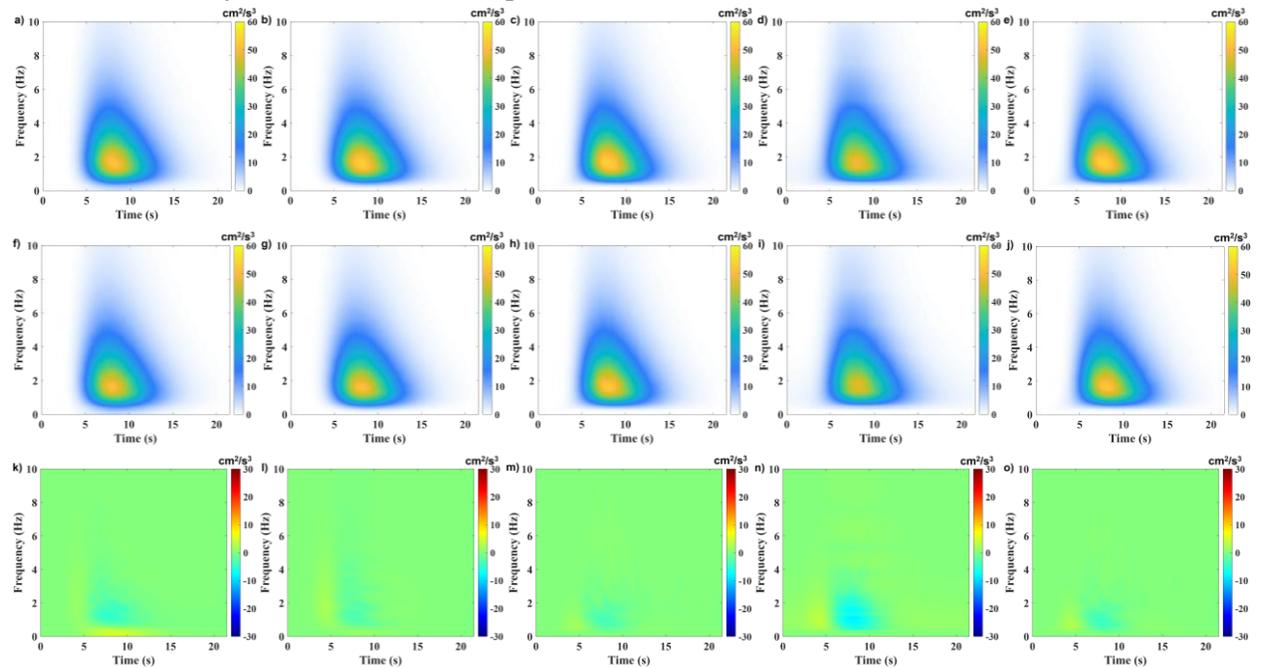

Figure 8. The first row shows the estimated mean of EPSD functions based on 10000 samples using different transforms (i.e., the STFT by using the box window with $h = 1$, the STFT by using the Gaussian window with $\sigma = 1$, the ST with $\kappa = 1$, the CWT by using HW with $(m, n) = (1, 2^{0.5})$, $c_0 = 0.01$, and $s_0 = 2^{0.5}$, and CWT by using GMW with $\beta = 20$, $\gamma = 3$, $c_0 = 0.01$, and $s_0 = 2^{0.1}$); the second row of plots shows the standard deviation of the estimated EPSD function, and the third row shows the difference between the estimated mean EPSD and its target (i.e., the first row – the target shown in Figure 7a). From the left to the right, the results shown are for the STFT with box

window, STFT with Gaussian window, ST, CWT with HW, and CWT with GMW.

To see the magnitude of the residuals for this considered example, we evaluate $C_{1,0}$, $C_{2,0}$, $C_{0,1}$, $C_{0,2}$, and $C_{0,0}$ (which are functions of time and frequency) by considering the target EPSD function shown in Eq. (59). The obtained residuals are shown in Figure 9 for each considered transform. The results presented in the figure indicate that the magnitude of the estimated residual depends on the employed transform; $R_S(f,\tau;2)$ is smaller than $R_{STFT}(f,\tau;2)$ with the box window or the Gaussian window. The magnitude of $R_W(f,\tau;1)$ obtained by applying the CWT with HW is greater than that obtained by applying the CWT with GMW. The relatively large value of $R_W(f,\tau;1)$ shown in Figure 9 is due to the large value of $r_W(0,1)$ at low frequencies (see Figure 5). However, it must be emphasized that the estimated residual depends on the selected parameters for the considered transforms. Therefore, the observations should not be generalized. In addition, the residuals with higher-order derivatives are not evaluated in the present study.

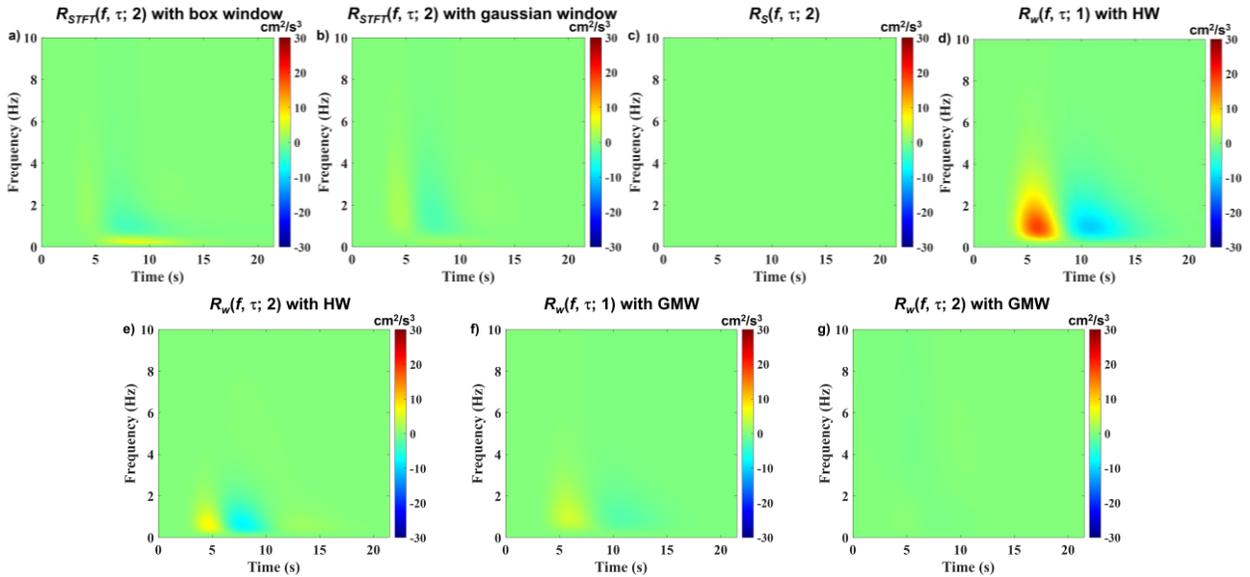

Figure 9. Calculated residuals according to the equations summarized in Table 1 for the given PSD function presented in Eq. (59).

### 4.2 Comparison of the estimated EPSD function from actual seismic ground motion records

To illustrate the application of STFT, ST, and CWT in estimating the EPSD function for actual seismic ground motion records, we consider a set of 12 ground motion record components oriented in the E-W direction for a seismic event that occurred on January 16, 1986. The event has a local magnitude of 6.1, a focal depth of 10.2 km, and an epicentral distance of 25.2 km. The records were obtained by the Large Scale Seismic Test (LSST) array in Lotung, Taiwan, where the separation between any two recording sites is less than 100 m (http://www.earth.sinica.edu.tw/). The arrangement of the recording site and one of the considered record components are shown in Figures 10a and 10b  By considering the five different transforms with the parameters that are the same as those used for results presented in Figure 8 and the 12 record components, the calculated means of the EPSD function are shown in Figures 10c to 10g for each considered transform. As can be observed from these plots, the estimated mean of EPSD by applying different transforms is relatively consistent, especially for those obtained by using the STFT with Gaussian window, ST, and CWT with GMW. Using the mean of the EPSD function that is obtained by applying STFT with the Gaussian window as the reference, the differences in the estimated mean of EPSD function by using different transforms are shown in Figures 10h to 10k. Figure 10h indicates that the use of STFT with the box window or Gaussian window can lead to different estimated EPSD. Figure 10j indicates that significant over- and underestimation of the EPSD, as compared to that obtained by using STFT with the Gaussian window, can result at a few locations within the time-frequency domain if the CWT with HW is employed. Figures 10i and 10k exhibit similar patterns, indicating that the use of the ST and CWT with the GMW results in a similar estimate of the EPSD function. It must be emphasized that the observed differences must be interpreted within the context of the selected parameters for the transforms, which are given in Figure 8.

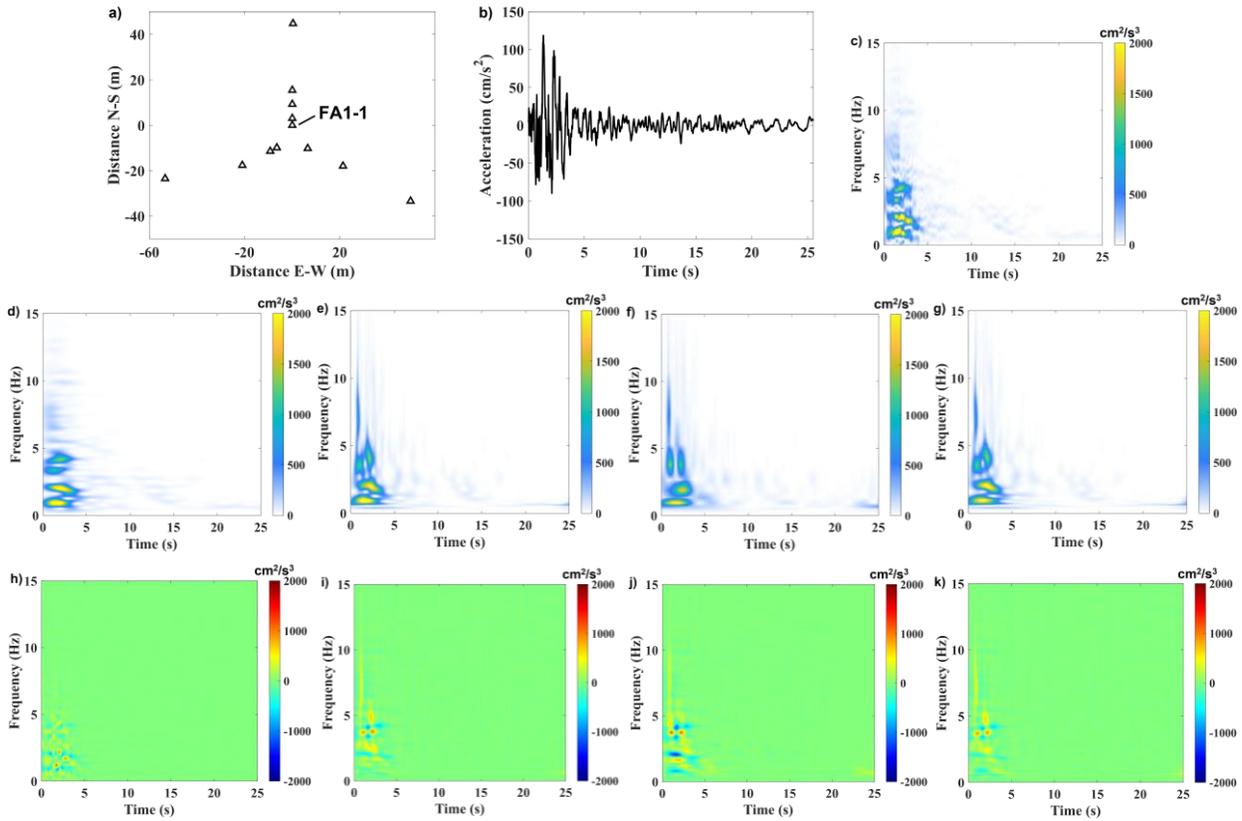

Figure 10. The LSST array and the recorded component at FA1–1 station are shown in a) and b). The estimated means of the EPSD by using different transforms are shown in c) to g) for STFT with box window, STFT with Gaussian window, ST, CWT with HW, and CWT with GMW, respectively. The difference between the estimated EPSD by using one transform (STFT with box window, ST, CWT with HW, and CWT with GMW) minus that obtained by using ST with Gaussian window is shown in h) to k), respectively.

### 4.3 Comparison of the estimated EPSD function from actual thunderstorm wind record

In this section, we show the estimated EPSD function for wind speed recorded from an array of anemometers located on a tower. The recorded winds were for the rear-flank downdraft, near Lubbock, Texas on June 4, 2002, with the weather condition leading to the recording described in Orwig and Schroeder (2007). The recorded winds obtained from anemometers located at 10 and 15 m height on Tower 4 are shown in Figures 11a and 11b.

For the numerical analysis, we first remove the time-invariant constant mean wind speed from the recorded wind speeds for each of the considered records. We calculated the EPSD function for the zero-mean wind speed records by applying different transforms discussed in the previous section and summarized in Table 1. The estimated PSD functions for the considered transforms are presented in Figures 11c to 11g for the wind recorded at 10 m height and in Figures 11h to 11l for the wind recorded at 15 m height. For the analysis, $h = 50$ is considered for the STFT with the box window; $\sigma = 50$ is considered for the STFT with the Gaussian window; $\kappa = 1$ is used for ST; $(m, n) = (1, 2^{0.5})$, $c_0 = 0.5$, and $s_0 = 2^{0.5}$, are used for the CWT with HW, and $\beta = 20$, $\gamma = 3$, $c_0 = 0.5$, and $s_0 = 2^{0.1}$ are used for the CWT with GMW. Note that $h = 50$ and $\sigma = 50$ (rather than $h = 1$ and $\sigma = 1$ which are used for processing the ground motion records) are employed to capture the energy distribution at the low frequencies. A sensitivity analysis is also carried out by varying $h$ and $\sigma$. The obtained results (that are not shown) indicate that a very poor time-frequency representation of the energy at low or high frequencies is obtained as the values of $h$ and $\sigma$ used for the analysis are very small or very large, respectively. Even with $h = 50$ and $\sigma = 50$, the resolution of the energy at low frequencies is not consistent with that obtained by using the remaining three transforms. In general, the EPSD function estimated by using ST agrees well with that obtained by using the CWT with GMW. The use of the CWT with HW results in relatively blurred energy concentrations at high frequencies. We emphasize that the observed differences must be interpreted within the context of the selected parameters for the transforms used for the analysis.

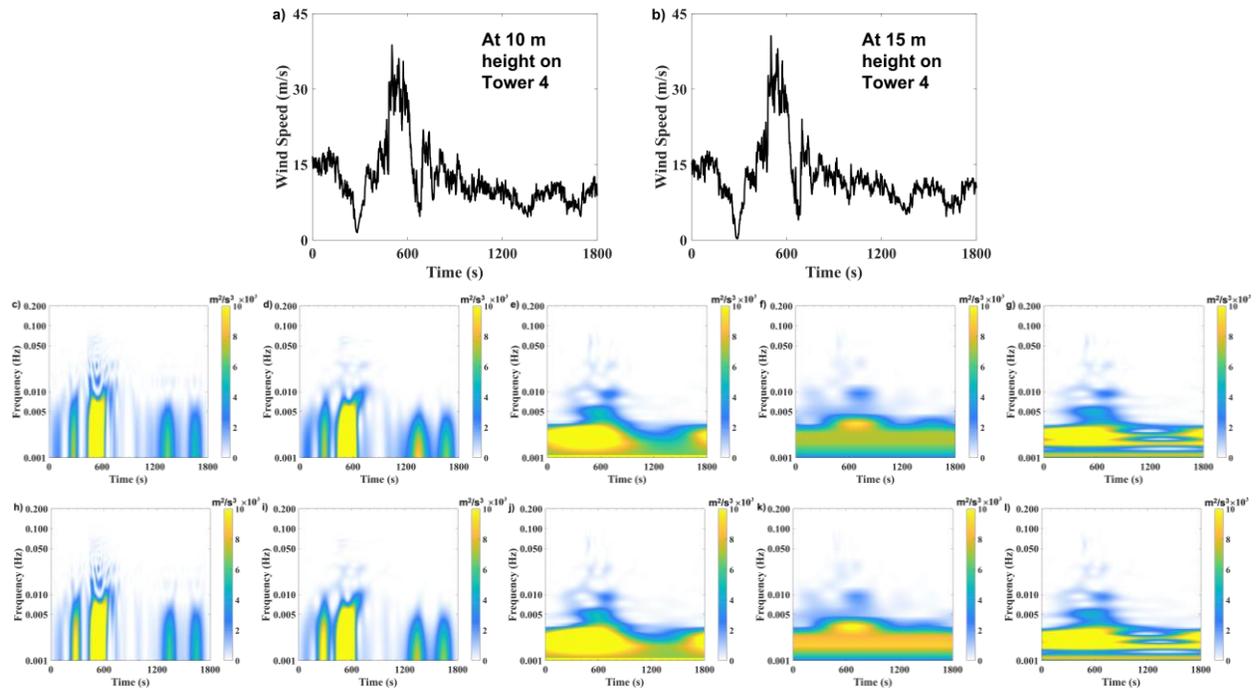

Figure 11. Time histories of downburst records are shown in a) and b); the calculated EPSD functions are shown in d) to g) for the wind record at 10 m height but with the time-invariant mean wind speed removed, and the calculated EPSD functions are shown in h) to l) for the wind record at 15 m height but with the time-invariant mean wind speed removed.

**5.0 Summary and conclusions**

A key aspect of using the evolutionary process to model and simulate an evolutionary process is to estimate the evolutionary power spectral density (EPSD) function from available samples. There are two popular approaches - one based on the short-time Fourier transform (STFT), and the other based on continuous wavelet transform (CWT)- for estimating EPSD function in the literature. Both rely on the concept of slowly varying modulation function, although the quantification of the effect of the ''slow'' variation in the estimated EPSD is elusive.

In the present study, simple-to-use formulas for estimating EPSD function are derived and presented based on the STFT, S-transform (ST), and CWT. Equations for calculating the residual for the estimated EPSD function are also derived by considering these transforms. The relation between the estimated EPSD function and the power distribution in the transform domain for each of the selected transforms is established. The relation indicates that the power distribution in the transform domain can be used directly or can be scaled and used as an estimate of the EPSD function (see Table 1).

In addition, it is shown that if the ''slow'' variation means that the second or higher (crossed or non-crossed) order derivatives of the EPSD function are equal to zero, the use of STFT (i.e., but without further smoothing in time) can still be associated with a residual term that depends on the multiplication of two first order derivatives and the adopted window. In general, second and higher-order derivatives all contribute to the residual. That is, the derived equations for the residual make the slow variation quantifiable in the context of estimating the EPSD function.

**Data availability statement**

Some or all data used during the study are available from the corresponding author by request.
**Acknowledgments**
Financial support was received from the Natural Sciences and Engineering Research Council of Canada (RGPIN-2016-04814, for HPH). I am grateful to X.Z. Cui for many fruitful discussions during this study and to Y.X. Liu for preparing some of the figures.